\pdfoutput=1
\documentclass[aps,twocolumn, floatfix, prx]{revtex4}
\usepackage{graphicx}
\DeclareGraphicsExtensions{.pdf}
\usepackage{amsmath,amssymb,
bbold,
bm,color}
\usepackage{float}
\usepackage{epstopdf}

\newcommand{\bk}{{\bm k}}

\newcommand{\bp}{{\bm p}}
\newcommand{\br}{{\bm r}}
\newcommand{\bv}{{\bm v}}
\newcommand{\bR}{{\bm R}}
\newcommand{\bB}{{\bm B}}

\newcommand{\bG}{{\bm G}}

\newcommand{\bA}{{\bm A}}
\newcommand{\bS}{{\bm S}}

\newcommand{\bj}{{\bm j}}
\newcommand{\bsig}{{\bm \sigma}}
\newcommand{\btau}{{\bm \tau}}
\newcommand{\bOmega}{{\bm \Omega}}

\newcommand{\bee}{\begin{equation}}
\newcommand{\ee}{\end{equation}}

\newcommand{\ket}[1]{| #1 \rangle}

\begin{document}

\title{Strongly interacting Majorana
  fermions}

\author{Ching-Kai Chiu}
\author{D.I. Pikulin}
\author{M. Franz}
\affiliation{Department of Physics and Astronomy, University of
British Columbia, Vancouver, BC, Canada V6T 1Z1}
\affiliation{Quantum Matter Institute, University of British Columbia, Vancouver BC, Canada V6T 1Z4}

\begin{abstract}
Interesting phases of quantum matter often arise when the constituent particles -- electrons in solids -- interact strongly. Such strongly interacting systems are however quite rare and occur only in extreme environments of low spatial dimension, low temperatures or intense magnetic fields. Here we introduce a new system in which the fundamental electrons interact only weakly but the low energy effective theory is described by strongly interacting Majorana fermions. The system consists of an Abrikosov vortex lattice in the surface of a strong topological insulator and is accessible experimentally using presently available technology. The simplest interactions between the Majorana degrees of freedom exhibit an unusual nonlocal structure that involves four distinct Majorana sites. We formulate simple lattice models with this type of interaction and find exact solutions in certain physically relevant one- and two-dimensional geometries. In other 
cases we show how our construction allows for the experimental realization of interesting spin models previously only theoretically contemplated.

\end{abstract}

\date{\today}

\maketitle

\section{Introduction}

When fermions partially occupy a band that is flat their kinetic energy
is quenched and interactions, even when nominally weak, can have a
profound effect on the  ground state of the system. This
paradigm is realized, with spectacular results, in 2D electron gases
in perpendicular  magnetic field where the interplay between the  flat
Landau level band structure and the Coulomb interaction gives
rise to fractional quantum Hall effect (FQHE) with all its remarkable
phenomenology \cite{tsui1,laughlin1}. More recently, it has been realized that magnetic field
is not necessary for the formation of FQHE states: one can
obtain these, at least in principle, from lattice models that are
tuned  so that their conduction band is (nearly) flat and at the same
time exhibits a non-zero Chern number making it topologically
nontrivial \cite{tang,sun,neupert,hu,sheng1,regnault1} .  When these conditions are met
one can achieve FQHE without magnetic field and there has been considerable
interest in such systems recently. 
In practice, however, it is not clear how a lattice system with a
topologically non-trivial flat band could be realized experimentally
because the occurrence of a flat band typically requires considerable fine
tuning of the overlap integrals which are given in solids by crystal
chemistry and this is, in most cases,  not continuously
tunable. Proposals exist  to artificially engineer such systems in
optical lattices and dipolar spin systems \cite{cooper1,lukin1}.

In this study we introduce a physical lattice system in which a
completely flat band can be obtained by tuning a single parameter. The band is unusual because its fundamental  degrees of freedom are Majorana
fermions \cite{alicea_rev,beenakker_rev,stanescu_rev,elliot_rev}. In
the flat band regime the Hamiltonian is dominated by the interaction
term and the system is therefore inherently strongly
correlated. Its phenomenology
differs substantially from the FQHE paradigm. Interesting phases
nevertheless arise and we explore them  in some detail.

\begin{figure}[t]
\includegraphics[width = 8.0cm]{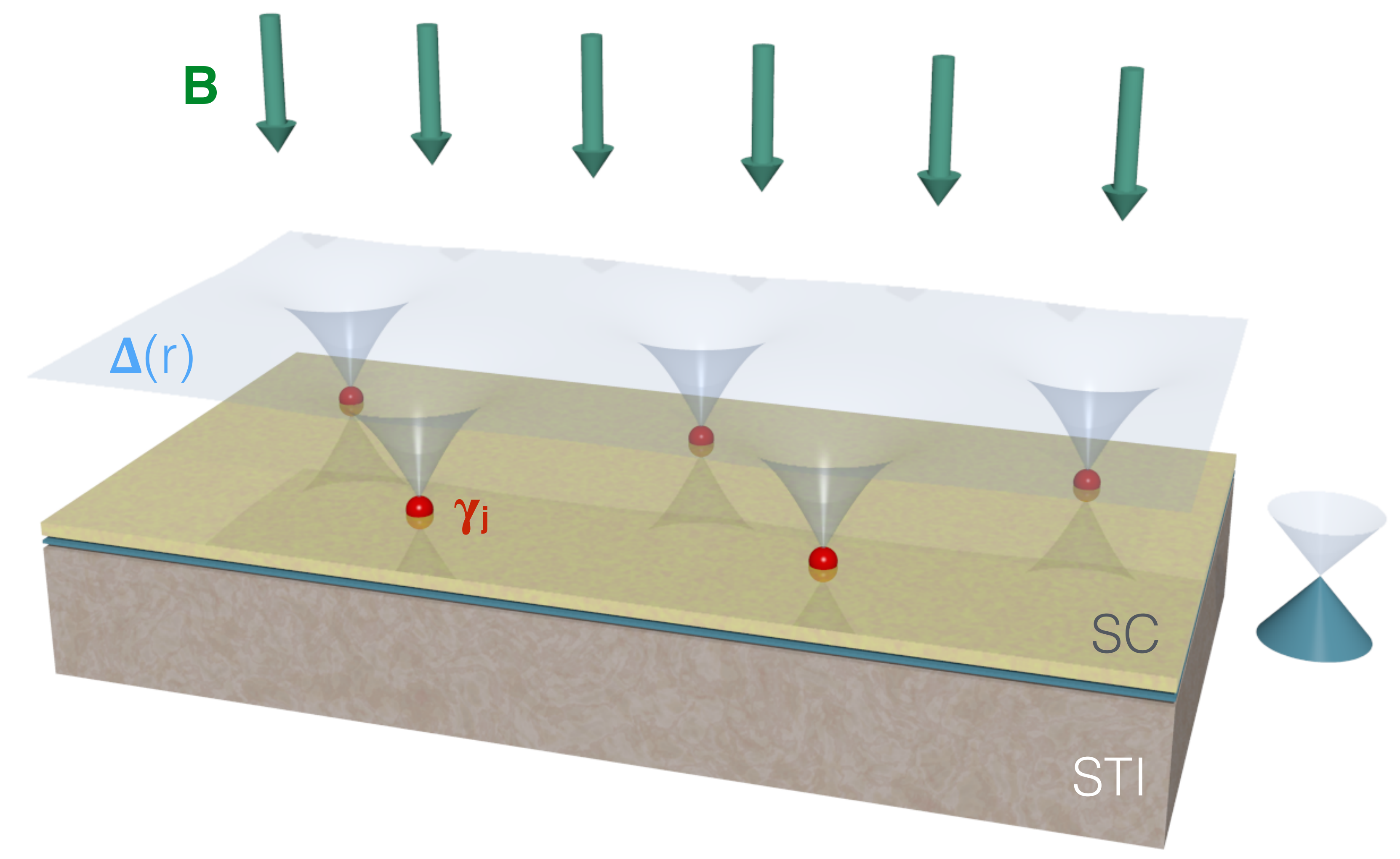}
\caption{{ Schematic depiction of the system based on Fu-Kane model \cite{fu1}}. Superconducting order is induced in the surface of a strong topological insulator (STI) gapping out the protected surface states with the Dirac dispersion. Magnetic field ${\bf B}$ is then applied to induce Abrikosov vortices in the SC order parameter $\Delta(\br)$. Each vortex hosts an unpaired Majorana zero mode $\gamma_\bj$.   
}\label{fig1}
\end{figure}

The specific system we consider is depicted in Fig.\ \ref{fig1} and consists of an Abrikosov lattice of vortices in
the surface state of a strong topological insulator (STI) that has been made
superconducting, either intrinsically as suggested by recent
experiments \cite{koren1,elbaum1}, or through the proximity effect with an
adjacent ordinary superconductor
\cite{sacepe1,qu1,williams1,cho1,xu0}. Theoretically, the situation is described by the Fu-Kane
model \cite{fu1}  which also famously predicts that each vortex in the SC
order parameter binds a Majorana zero mode. Tentative
experimental evidence for such zero modes has been recently reported
in Bi$_2$Te$_3$/NbSe$_2$ heterostructures \cite{xu1}. 

When two vortices are brought together their Majorana
wavefunctions start overlapping and, generically, the zero modes
split. In the vortex lattice one thus expects formation of a Majorana
band whose bandwidth increases as the lattice becomes denser. This
is indeed observed in analytical and numerical calculations 
\cite{galitsky1,galitsky2,kraus1,kou1,silaev1,biswas1,hung1}. However, as we discuss in more detail below, in the
special case when the chemical potential $\mu$ of
the STI coincides with the Dirac point of the surface state (hereafter
referred to as the neutrality point) the band formation can be avoided. This is because the Fu-Kane model at the neutrality exhibits an extra
``chiral'' symmetry and, as observed by Teo and Kane \cite{teo1},
vortex defects are then in topological class BDI described by an {\em integer} (as
opposed to Z$_2$ valued) invariant. Physically, this means that the
total number of exact zero modes in the system is equal to the total
vorticity, i.e.\ the total number $N_V$ of vortices present in the
system. This is to be contrasted with the Z$_2$ classification that
applies away from the neutrality point and implies $(N_V\! \mod 2)$ exact
zero modes. 

The above considerations imply that at the neutrality
point  the chiral symmetry  present in the Fu-Kane model prohibits Majorana zero modes from hybridizing, independent of their detailed geometric arrangement. The Majorana band that arises in the vortex
lattice therefore remains {\em completely flat}. In this situation one may expect interactions to play an important role in determining the collective quantum state of the system. In what follows we explore some of the interesting strongly correlated phases of Majorana fermions that arise in such vortex lattice models. We find that, remarkably, certain strongly interacting models of this type admit exact solutions owing to the presence of an  extensive number of conserved quantities. In other cases exact solutions are not available but the Hamiltonians can be mapped onto spin models, some of which have been studied previously and some that appear new.

In Sec.\ II below we review the general symmetry arguments that indicate the absence of the zero mode hybridization in the Fu-Kane model at neutrality in greater detail. We then outline how this physics arises in a concrete model calculation and use this model in Sec.\ III to derive the form of the interaction terms and estimate their strength, as well as discuss the effects of small detuning from neutrality on the effective low-energy Hamiltonian of the system. In Sec.\ IV we proceed to analyze various interacting lattice models with Majorana fermions that can arise in vortex lattices in different one- and two-dimensional geometries. We conclude in Sec.\ V by discussing prospects for experimental realization and observation of these lattice models in physical systems and we speculate about some novel phases of strongly interacting Majorana matter that can be potentially engineered with help of the tools introduced in this study.

\section{Majorana flat bands in vortex lattices}

\subsection{Zero modes in Fu-Kane model} \label{zero modes}

Fu and Kane \cite{fu1} envisioned inducing superconductivity in the surface state
of a 3D topological insulator by covering it in with a thin film of an
ordinary $s$-wave superconductor such as Pb or Nb. Alternately, superconductivity could appear as an intrinsic instability of the surface state \cite{koren1,elbaum1} or be induced in thin STI flakes through their bulk by placing them on a SC substrate \cite{xu1}. In either case the 
second-quantized Hamiltonian  describing such superconducting STI surface state  can be written as
\begin{equation}\label{abel1}
{\cal H}=\int d^2r \hat{\Psi}^\dagger_\br H_{\rm FK}(\br)\hat{\Psi}_\br,
\end{equation}
where
$\hat\Psi_\br=(c_{\uparrow\br},c_{\downarrow\br},c^\dagger_{\downarrow\br},-c^\dagger_{\uparrow\br})^T$
is the Nambu spinor and 
\begin{equation}\label{abel3}
H_{\rm FK}(\br)= 
\begin{pmatrix}
-\mu & v p_- & \Delta(\br) & 0 \\
v p_+ & -\mu & 0 & \Delta(\br) \\
\Delta^*(\br) & 0 & \mu & -v p_- \\
0 & \Delta^*(\br) & -v p_+ & \mu
\end{pmatrix},
\end{equation}
with $p_{\pm}=p_x\pm ip_y$ and  $\mu$  the
chemical potential. The diagonal $2\times 2$ blocks describe the kinetic energy of
the STI surface state (single Dirac fermion with velocity $v$) while the off-diagonal blocks encode the SC pair potential.

As the first step we are interested in finding the eigenstates $\Phi_n(\br)$ of $H_{\rm FK}(\br)$ in
the presence of a single Abrikosov vortex. For a vortex placed at the origin we write
\begin{equation}\label{abel4}
\Delta(\br)=\Delta_0(r)e^{-i(n\varphi+\theta)},
\end{equation}
where $\Delta_0(r)$ is a real function of the distance, $\varphi$
represents the polar angle and $\theta$ denotes an arbitrary constant
phase offset due to other vortices that could be present in the system
far away from the origin. Integer $n$ denotes the vorticity. 
Single-valuedness of the Hamiltonian
dictates that  $\Delta_0(r)$ vanishes at the origin. Energy
considerations further show that  $\Delta_0(r)\sim r^{|n|}$ for small $r$.

To find the zero modes of  $H_{\rm FK}(\br)$ in the presence of a
vortex it is useful to first perform a unitary transformation
$\tilde{H}_{\rm FK}=UH_{\rm FK}U^{-1}$ with  
\begin{equation}\label{abel5}
U=
\begin{pmatrix}
1 & 0 & 0 & 0 \\
0 & 0 & 0 & 1 \\
0 & 0 & 1 & 0 \\
0 & 1 & 0 & 0
\end{pmatrix},
\end{equation}
which brings the Hamiltonian into the following form
\begin{equation}\label{abel6}
\tilde{H}_{\rm FK}= 
\begin{pmatrix}
M & D \\
D^\dagger & -M 
\end{pmatrix}, \ \ \ \ 
D= 
\begin{pmatrix}
\Delta(\br) & p_- \\
-p_+ & \Delta^*(\br) 
\end{pmatrix}.
\end{equation}
and $M={\rm diag}(-\mu,\mu)$.
The transformed Hamiltonian acts on the modified Nambu spinor
$\hat\Psi_\br=(c_{\uparrow\br},-c^\dagger_{\uparrow\br},c^\dagger_{\downarrow\br},c_{\downarrow\br})^T$. 
Passing into the polar coordinates and making use of the identity $p_\pm=e^{\pm i\varphi}(-i\partial_r\pm r^{-1}\partial_\varphi)$ we may write 
\begin{equation}\label{abel8}
D= 
\begin{pmatrix}
e^{-i(n\varphi+\theta)}\Delta_0(r) & e^{-i\varphi}(-i\partial_r-{\partial_\varphi\over r}) \\
-e^{i\varphi}(-i\partial_r+{\partial_\varphi\over r}) & e^{i(n\varphi+\theta)}\Delta_0(r)
\end{pmatrix},
\end{equation}
where we have set $v=\hbar=1$.
We now temporarily focus on the neutrality point where $M=0$ and the
Hamiltonian (\ref{abel6})  is purely off-diagonal.
When looking for the zero modes the off-diagonal form has a distinct advantage: the zero modes necessarily have the spinor structure $(\psi(\br),0)^T$ and $(0, \chi(\br))^T$ where $\psi(\br)$ and $\chi(\br)$ are two-component zero modes of $D^\dagger$ and $D$ respectively.
For a singly quantized vortex $(n=1)$ it is easy to show that there exists a normalizable zero mode of $D$ of the form
\begin{equation}\label{abel9}
\chi_0(\br)={1\over\sqrt{2}}
\begin{pmatrix}
e^{-i(\theta/2-\pi/4)}\\
e^{i(\theta/2-\pi/4)}
\end{pmatrix}
f_0(r), 
\end{equation}
with  
\begin{equation}\label{abel99}
f_0(r)=Ae^{-\int_0^r\Delta_0(r')dr'},
\end{equation}
while $D^\dagger$ does not have a normalizable zero mode. The field operator of the zero mode reads
\begin{equation}\label{abel10}
\gamma={1\over\sqrt{2}}\int d^2r \left[e^{i(\theta/2-\pi/4)}c_{\br\downarrow}+e^{-i(\theta/2-\pi/4)}c^\dagger_{\br\downarrow}\right]f_0(r).
\end{equation}
As expected, the zero mode represents a Majorana particle,
$\gamma^\dagger=\gamma$. 
For $\mu\neq 0$ the structure of the zero
mode wavefunction becomes slightly more complicated \cite{galitsky2}; in addition to the 
exponential decay it exhibits an oscillatory behavior $\sim\sin{kr}$
where $k$ is a wavevector close to the Fermi  wavevector $k_F=\mu/v$. 

When multiple well-separated vortices are present in the system then each will harbor a Majorana zero mode. Their respective creation operators $\gamma_j$ satisfy the anticommutation algebra \cite{alicea_rev,beenakker_rev,stanescu_rev,elliot_rev}
\begin{equation}\label{can2}
\{\gamma_{i},\gamma_{j}\}=2\delta_{ij},
\ \ \ \gamma_{i}^\dagger=\gamma_{i},
\end{equation}
characteristic of Majorana fermions. The latter follows directly from Eq.\ (\ref{abel10}) generalized to multiple vortices  and the canonical anticommutation relations for the electron operators $c_{\br\sigma}$. The expected non-Abelian exchange statistics of vortices containing Majorana zero modes \cite{read1,ivanov1} becomes apparent when one considers adiabatic exchange of two such vortices. In what follows we shall deal with vortices pinned at fixed positions and their non-Abelian properties will therefore not play an essential role in our considerations.

\subsection{Symmetry considerations}
In the presence of multiple vortices that are closely spaced the fate of the zero modes associated with a single isolated vortex will depend on the symmetries of the underlying Hamiltonian, as discussed in detail by Teo and Kane \cite{teo1}. We now briefly review their analysis as relevant to the Hamiltonian (\ref{abel3}). To facilitate the discussion we rewrite the latter in a more compact notation 
\begin{equation}\label{sym1}
H_{\rm FK}= \tau^z(\bp\cdot\bsig-\mu)+\tau^x\Delta_1+\tau^y\Delta_2
\end{equation}
where $\Delta=\Delta_1+i\Delta_2$ and $\bsig$, $\btau$ are Pauli matrices in spin and Nambu spaces, respectively. The Hamiltonian (\ref{sym1}) respects the particle-hole symmetry generated by $\Xi=\sigma^y\tau^y K$ 
($\Xi^2=+1$, $K$ denotes complex conjugation) and, for a purely real gap function $\Delta$, also the physical time reversal symmetry $\Theta=i\sigma^y K$ ($\Theta^2=-1$). In the presence of vortices $\Delta$ becomes complex and the time reversal symmetry is broken. Fu-Kane model with vortices therefore defines symmetry class D in the Altland-Zirnbauer classification which according to Ref.\ \onlinecite{teo1} implies a Z$_2$ classification for the zero modes associated with point defects such as vortices. Physically, this means that a system with total vorticity $N_V$ will have $(N_V\! \mod 2)$ exact zero modes, in accord with the expectation that any even number of Majorana zero modes will generically hybridize and form  complex fermions at non-zero energies.

However, in the special case when $\mu=0$, Hamiltonian (\ref{sym1}) respects a fictitious time reversal symmetry with $\tilde{\Theta}=\sigma^x\tau^x K$ ($\tilde{\Theta}^2=+1$), even in the presence of vortices. At the neutrality point, the two symmetries $\Xi$ and $\tilde{\Theta}$ define a BDI class with chiral symmetry $\Pi=\Xi\tilde{\Theta}=\sigma^z\tau^z$. This, according to  Ref.\ \onlinecite{teo1} implies an integer classification of zero modes associated with point defects. A system with total vorticity $N_V$ will thus exhibit $N_V$ exact zero modes, irrespective of the details such as the geometric arrangement of the individual vortices. Below we illustrate how this interesting behavior emerges in a concrete model calculation.

We remark that Fu-Kane model at the neutrality point coincides with
  the Jackiw-Rossi model \cite{rossi1} well known in particle physics, where the $\mu=0$ condition is enforced by Lorentz invariance.  An index
 theorem for Dirac fermions applied to this model \cite{weinberg1}  is known to connect the total vorticity
 with the number of protected fermionic zero modes. This property of
 the Fu-Kane model has been previously noted in Ref.\ \cite{galitsky2}.

\subsection{Zero mode hybridization in a vortex lattice}

We now study the zero mode hybridization using the microscopic wavefunctions obtained above in subsection \ref{zero modes}. To begin
consider two vortices located at points $\bR_1$ and $\bR_2$, such
that $|\bR_1-\bR_2|\gg \xi$. The two-vortex Hamiltonian $H_{\rm FK}^{(2)}$
still has the structure displayed in Eq.\ (\ref{abel6}) except that $\Delta(\br)$
now encodes vortices at $\bR_1$ and $\bR_2$.
We can seek its low-energy eigenstates in the
basis spanned by the  zero mode wavefunctions
$\Phi_1(\br)=(0,\chi_0(\br-\bR_1))^T$ and
$\Phi_2(\br)=(0,\chi_0(\br-\bR_2))^T$. If we denote the two
Majorana operators as $\gamma_1$ and $\gamma_2$ then the zero mode splitting
comes from the term $it_{12}\gamma_1\gamma_2$ with  the overlap integral $it_{12}=\langle
\Phi_1|H_{\rm FK}^{(2)}|\Phi_2\rangle $. At the neutrality point the
matrix element $t_{12}$ trivially evaluates to zero because $|\Phi_1\rangle$ is
orthogonal to $H_{\rm FK}^{(2)}|\Phi_2\rangle $ for arbitrary positions
$\bR_1$ and $\bR_2$. The zero modes therefore remain exact as expected on the basis of the symmetry argument presented above. 

Away from the
neutrality point we find\cite{biswas1}
\begin{eqnarray}\label{abel11}
it_{12}&=&\int d^2r \chi_0^\dagger(\br-\bR_1)(-M)\chi_0(\br-\bR_2)
\nonumber \\
& =&
i\mu\sin{\left({\theta_1-\theta_2\over 2}\right)} F_{12} 
\end{eqnarray}
with $F_{12}=\int d^2r f_0(\br-\bR_1)f_0(\br-\bR_2)$; the
overlap is proportional to $\mu$ and is generally nonzero.

If there are many vortices in the system then the overlap integrals
remain zero at the neutrality point and are given by a generalization
of Eq.\ (\ref{abel11}) when $\mu\neq 0$. A system of many vortices in
a superconductor (or a  charged superfluid) is only stable in the
presence of an externally applied magnetic field $\bB$ \cite{tinkham1}.  
To describe a
realistic vortex lattice we must therefore include magnetic field by
performing a minimal substitution $\bp\to\bp-\tau^z(e/c)\bA$ in the Hamiltonian
(\ref{sym1}). One can show that the presence of $\bA$
does not qualitatively change the  zero mode wavefunction (\ref{abel9}) associated with an individual vortex. However,
the phase difference $(\theta_1-\theta_2)/2$ in the overlap integral Eq.\ (\ref{abel11}) must be replaced by its gauge invariant generalization
\begin{equation}\label{abel112}
\omega_{12} =\int_{\br_1}^{\br_2}\left({1\over 2}\nabla\theta-{e\over
    c}\bA\right)\cdot d{\bf l},
\end{equation}
where the integral is taken along the straight line between vortex
positions $\br_1$ and $\br_2$. This result can be obtained by an explicit calculation but also follows from a simple general argument: because the overlap amplitudes
$|t_{ij}|$ are potentially measurable  physical quantities they
cannot depend on an arbitrarily chosen gauge.  Some details of how one
evaluates the gauge invariant phases (\ref{abel112}) in the vortex lattice geometry are provided in Appendix \ref{phases}.

The low-energy effective Hamiltonian describing the Majorana zero
modes in a vortex lattice  can thus be written as 
\begin{equation}\label{kin1}
{\cal H}_{\rm kin} =\sum_{i,j} t_{ij} s_{ij} \gamma_i\gamma_j.
\end{equation}
Here we use a notation introduced in Ref.\ \onlinecite{grosfeld1}
where $t_{ij}$ is a real symmetric matrix representing the hopping
strength while $s_{ij}=e^{i\phi_{ij}}=\pm i$ are the Z$_2$ gauge
factors. The sign ambiguity arises from the fact that one can perform a local
Z$_2$ gauge transformation $\gamma_j\to -\gamma_j$ without affecting
the zero mode commutation algebra  (\ref{can2}). A product of $s_{ij}$
factors along a closed trajectory is however gauge invariant and
physically observable. It represents a Z$_2$ gauge flux and should be thought of as analogous to the magnetic flux expressed through Peierls factors in lattice models of charged particles. In the vortex lattice for a general polygon
formed by $n$ vortices the total phase is given by
\cite{grosfeld1}
\begin{equation}\label{kin2}
\sum_{\rm polygon}\phi_{ij}={\pi\over 2}(n-2).
\end{equation}
In the context of Eqs.\ (\ref{abel11},\ref{abel112}) the Z$_2$ gauge factors arise from the  fact that half of the phase difference enters the overlap integral (\ref{abel11}) and the $\sin{\omega_{ij}}$ function is thus not single valued in the presence of vortices. The physics of the associated branch cuts and how they give rise to the Z$_2$ gauge factors is further explained in
Appendix \ref{phases}.

We note that according to Eq.\ (\ref{kin2}) for both triangular and
square vortex lattices if $t_{ij}$ are non-zero, Majorana fermions
move in a background of non-trivial Z$_2$ flux. This makes even the
non-interacting problem interesting and leads to the rich physics of
``nucleated'' topological phases, explored in previous studies
\cite{lahtinen1,lahtinen2}.

As already noted, for $\mu\neq 0$ the Majorana wavefunctions exhibit
Friedel-like oscillations with lengthscale set by $k_F=\mu/v$. When intervortex distance $d$
is such that $k_F d\gtrsim 1$ then this leads to an oscillatory
behavior of the overlaps $t_{ij}$ with the distance. Such oscillations
in combination with disorder in vortex positions have been studied and
shown to produce interesting effects \cite{kraus1,lauman1}. In this
study we focus on the regime $k_F^{-1}>d\gtrsim\xi$ where the
oscillatory behavior can be neglected. Oscillations in this regime
have no effect on the hoppings
between near neighbors and are damped out by the exponential decay of
the wavefunctions on longer distances. As we will show in the next
Section it is precisely this regime where the interactions tend to
dominate over the kinetic energy and this is also where our interest lies.

\section{Interaction effects} \label{interaction}

\subsection{General considerations}

We showed in the previous Section that by tuning a single system parameter in the Fu-Kane model (the global chemical potential $\mu$) one can eliminate the hybridization between the Majorana zero modes bound to individual vortices. We demonstrated how this occurs in a specific microscopic model but we emphasize that this effect only depends on the system symmetries and not on the microscopic details. 

At the neutrality point, therefore, the Majorana band associated with
an arbitrary vortex lattice will be completely flat and the manybody
ground state will exhibit $2^{N_V/2-1}$-fold degeneracy under the conservation of fermionic parity. At the
non-interacting level this degeneracy is robust to any
symmetry-preserving perturbation. A question that naturally arises is
what physical effects (if any) are likely to remove this extensive
ground state degeneracy in a physical system. There are essentially
two possibilities: (i) symmetry breaking disorder and (ii)
interactions. It is clear that local fluctuations in the chemical
potential $\mu$, if allowed, will generate random hoppings $t_{ij}$
between nearby Majorana zero modes  and these will in turn remove the
ground state degeneracy. This is because non-zero fluctuating $\mu$
breaks the fictitious time reversal symmetry $\tilde{\Theta}$ of the
Fu-Kane Hamiltonian at neutrality returning its zero mode
classification back to class D. Majorana fermion systems with random
hoppings have been previously considered in a number of studies
\cite{kraus1,vasuda1,lauman1,lauman2} . 

In this work we focus on the
interactions whose effects are much less well understood. Accordingly, we shall
consider  systems in which  the interaction strength is much larger
than any perturbation arising from the disorder effects. We will 
show that conditions under which such an assumption can be
justified can indeed occur in physical systems.
Specifically, we consider  four-fermion terms that arise from
Coulomb or possibly other interactions present in the underlying solid
state system. Such interactions are generated even when both particle-hole symmetry $\Xi$ and the fictitious time reversal symmetry $\tilde{\Theta}$ are respected. This allows for a genuinely strongly correlated regime in which the physics is completely dominated by interactions and the kinetic energy is quenched.

Under these assumptions the leading perturbation to the
degenerate manifold of Majorana zero modes will arise from
electron-electron interactions that are necessarily present in the
underlying solid. If we denote by $\gamma_j$ the annihilation
operator of the Majorana zero mode belonging to the $j$-th vortex then
the simplest interaction term that can be constructed has the form
\begin{equation}\label{int1}
{\cal H}_{\rm int}=\sum_{ijkl}g_{ijkl}\gamma_i\gamma_j \gamma_k\gamma_l,
\end{equation}
where $g_{ijkl}$ are real constants representing the  interaction
strength. The reality of  $g_{ijkl}$ follows from the requirement
that  ${\cal H}_{\rm int}$ be hermitian. Furthermore, since the Majorana operators 
obey the  anticommutation algebra (\ref{can2})
only the part of $g_{ijkl}$ that is antisymmetric in all indices
contributes to ${\cal H}_{\rm int}$.
We note specifically that according to Eq.\ (\ref{can2})
$\gamma^\dagger_i\gamma_i=\gamma_i\gamma_i=1$ and the terms in
${\cal H}_{\rm int}$ with two identical indices reduce to fermion
hoppings, e.g.\ $g_{iikl}\gamma_i\gamma_i
\gamma_k\gamma_l=g_{iikl}\gamma_k\gamma_l$. However, such terms are not hermitian and one can show
that since $g_{iikl}=g_{iilk}$ they identically
vanish. The simplest interaction term thus involves Majoranas located
at four distinct vortices. Such a non-local interaction may be expected
to give rise to unusual physical properties. 
     
The expression in  Eq.\ (\ref{int1}) is cumbersome because for every
group of four vortices it contains 24 distinct permutations of the
$\gamma$ operators. It is thus preferable to rewrite  ${\cal H}_{\rm int}$ 
as a sum over all distinct groups of four vortices in each of which we
define a specific ordering of $\gamma$'s.  For example for the group
($\gamma_1$, $\gamma_2$, $\gamma_3$, $\gamma_4$) we write the
interaction term as 
\begin{equation}\label{int2a}
{\cal H}_{\rm int} ^{1234}=g \gamma_1\gamma_2 \gamma_3\gamma_4
\end{equation}
and
similarly for other groups with $\gamma$'s organized in order of increasing index $j$. The interaction term ${\cal H}_{\rm int} ^{1234}$ is allowed to introduce in the Hamiltonian since ${\cal H}_{\rm int} ^{1234}$ in Majorana operator basis automatically preserves particle-hole symmetry and is invariant under time reversal operation $\gamma_j \rightarrow \gamma_j$ and $i \rightarrow -i$.

In the next subsection we shall discuss the microscopic origin and the strength of coupling constants $g$. As we shall see the coupling strength
$g$  depends on the zero mode wavefunction
overlaps as well as the detailed form of the interaction potential
$V(\br)$. For our present purposes it will suffice to note that since the Majorana wavefunctions decay exponentially outside
the vortex core, the largest $g$ will occur for those groups of four
vortices that are packed closest together. In the following we shall often 
consider examples of lattice systems in which we retain
only such dominant interactions and neglect all $g$'s associated with
groups of vortices that are more spread out since they are smaller by
factors $\sim e^{-d/\xi}$ where $d$ is the intervortex distance and
$\xi$ the SC coherence length. For instance in the square vortex
lattice we shall retain $g_\square^{~}$ associated with an elementary square
plaquette and neglect all other couplings.

\subsection{Microscopic origin of the interaction terms}

Suppose we have
solved the single-electron problem in the presence of $N$ vortices. We
thus have the complete set of eigenfunctions $\Phi_n(\br)$  and
eigenenergies $E_n$ of $H_{\rm FK}^{(N)}$.  The second
quantized Hamiltonian (\ref{abel1})  can then be written in a diagonal
form ${\cal H}=\sum_n' E_n\hat{\psi}^\dagger_n\hat{\psi}_n +E_g$ where 
\begin{equation}\label{abel12}
\hat{\psi}_n=\int d^2r \Phi_n^\dagger(\br)\hat{\Psi}_\br
\end{equation}
are the eigenmode operators. The sum over $n$ is restricted to the
positive energy eigenvalues and $E_g$ is a constant representing the ground
state energy. At the neutrality point, according to our preceding
discussion,  $N$ of the $\hat{\psi}_n$ eigenmodes coincide with the  exact zero modes
belonging to the individual vortex cores. We denote these
$\gamma_j$ with  $j=1\dots N$.

The Coulomb interaction, appropriately screened, can be written as 
\begin{equation}\label{bb2}
U={1\over 2}\int\int d^2r d^2r'\hat{\rho}(\br)V(\br-\br')\hat{\rho}(\br'),
\end{equation}
where $V(\br)$ is the interaction potential and
$\hat{\rho}(\br)=c^\dagger_{\sigma\br}c_{\sigma\br}$ is the electron charge density
operator.  The latter can be expressed in terms of modified Nambu
spinors as  $\hat{\rho}(\br)=\hat{\Psi}^\dagger_\br O_\rho \hat{\Psi}_\br$
with $O_\rho={1\over 2}{\rm diag}(1,-1,-1,1)$. Next, by exploiting the
completeness of the eigenstates $\Phi_n(\br)$ we can invert Eq.\
(\ref{abel12}) to obtain 
\begin{equation}\label{abel13}
\hat{\Psi}_\br=\sum_n\Phi_n(\br)\hat{\psi}_n
\end{equation}
and express the charge density in terms of the eigenmode operators as
\begin{equation}\label{abel14}
\hat{\rho}(\br)=\sum_{n,m}\left[\Phi^\dagger_n(\br)O_\rho
\Phi_m(\br)\right] \hat{\psi}^\dagger_n\hat{\psi}_m. 
\end{equation}
Substituting this result into Eq.\ (\ref{bb2}) and projecting onto the
zero mode subspace we arrive at the interaction between Majorana
modes of the form
\begin{equation}\label{bb3}
U_0={1\over 2}\sum_{ijkl} \gamma_i\gamma_j  \gamma_k\gamma_l\int\int d^2r d^2r'\rho_{ij}(\br)V(\br-\br')\rho_{kl}(\br'),
\end{equation}
where 
\begin{equation}\label{abel15}
\rho_{ij}(\br)=\left[\Phi^\dagger_i(\br)O_\rho\Phi_j(\br)\right]. 
\end{equation}
Comparing Eqs.\ (\ref{bb3}) and (\ref{int1}) we may read off an expression for $g_{ijkl}$, 
\begin{equation}\label{int3}
g_{ijkl}={1\over 2}\int\int dr^2 dr'^2\rho_{ij}(\br)V(\br-\br')\rho_{kl}(\br').
\end{equation}

At the neutrality point we can use Eq.\ (\ref{abel9}) to write
\begin{equation}\label{abel16}
\rho_{ij}(\br)=-{i\over 2}\sin{\left(\theta_i-\theta_j\over 2\right)} f_0(\br-\bR_i)  f_0(\br-\bR_j).
\end{equation}
In the vortex lattice when magnetic field is present the phase difference is to be replaced by $\omega_{ij}$ defined in Eq.\ (\ref{abel112}).
Noting the antisymmetry $\rho_{ij}(\br) =- \rho_{ji}(\br)$, the
expression for the interaction
parameter $g$ defined in Eq.\ (\ref{int2a})
for every such group of four vortices can be written as
\begin{equation}\label{int2}
g=\epsilon^{ijkl}g_{ijkl}
\end{equation}
where $\epsilon^{ijkl}$ is the totally antisymmetric tensor.
This can be further simplified, for
the group of four Majoranas $\gamma_1\dots\gamma_4$, as
\begin{equation}\label{abel17}
g=8(g_{1234}+g_{4123}-g_{1324}).
\end{equation}
The three distinct terms can now be evaluated with the help of Eq.\
(\ref{abel16}).

\subsection{Estimate of the interaction strength}

Since according to our analysis in Sec. II the hopping amplitudes between Majorana fermions $t_{ij}$ can be tuned to zero by adjusting the chemical potential the system will be in the strong interaction regime for any non-zero value of $g$. In practice, of course, we need $g$ sufficiently large to be able to observe the interaction effects in a sample with realistic levels of disorder and at nonzero temperature $T$. We thus require an estimate of $g$ relevant to a realistic situation. For concreteness, we consider the vortex lattice in the Bi$_2$Te$_3$/NbSe$_2$ heterostructure discussed in Ref.\ \onlinecite{xu1}.

\begin{figure*}[t]
\includegraphics[width = 16.0cm]{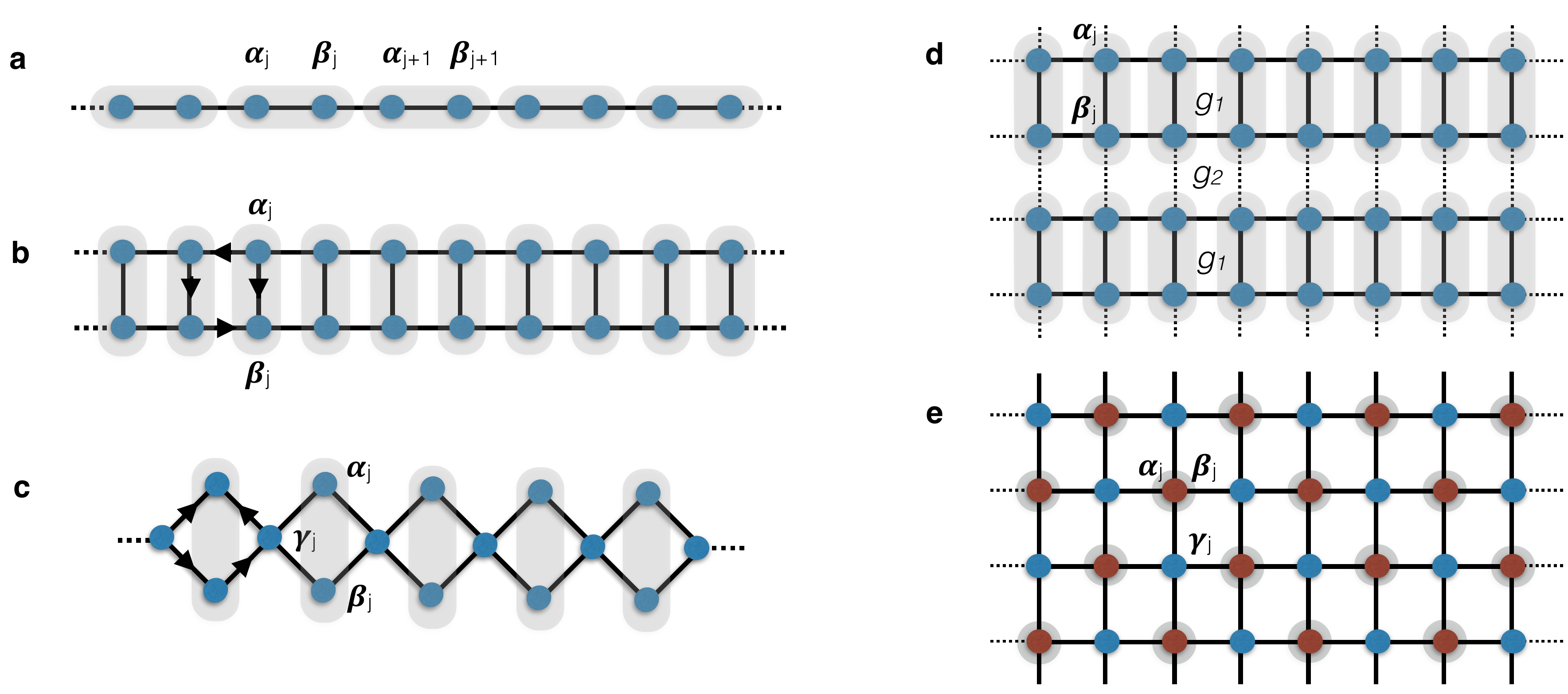}
\caption{ Lattice structures for models with strongly interacting Majorana fermions. {\bf a} Simple 1D chain, {\bf b} two-leg ladder and {\bf c} the diamond chain. {\bf d} Simple square lattice with two types of plaquettes characterized by interaction strength $g_1$ and $g_2$ and {\bf e} the modified square lattice with alternate sites (rendered in red) occupied by double vortices. The arrows in panels  {\bf b} and {\bf c} indicate our choice of the Z$_2$ gauge factors for the Majorana hopping terms consistent with Eqs.\ (\ref{kin1}) and (\ref{kin2}). 
}\label{fig2}
\end{figure*}
In this situation we expect the Coulomb interaction to be well screened so that it is essentially point-like on the scale set by the SC coherence length $\xi$, i.e. $V(\br)\simeq V_0\delta(\br)$. The expression (\ref{int3}) for the coupling constant simplifies, becoming
\begin{equation}\label{int4}
g_{ijkl}={1\over 2}V_0\int dr^2\rho_{ij}(\br)\rho_{kl}(\br).
\end{equation}
Evaluation of the coupling constant in this limit thus involves an estimate of $V_0$, calculation of the overlap integral implied by Eq.\ (\ref{int4}), and a determination of the geometric prefactors coming from the phases $\theta_i$ indicated in Eq.\ (\ref{abel16}).  We begin with the latter as the phases are determined purely by the vortex lattice geometry. We first consider an elementary square in an infinite periodic square vortex lattice. 
The phase difference structure is discussed in Appendix \ref{phases} and is consistent with the one obtained in \cite{grosfeld1}. We then  obtain  $g_\square=-2V_0F_{1234}$ where  
\begin{equation}\label{int5}
F_{1234}=\int d^2r\Pi_{j=1}^4f_0(\br-\bR_j).
\end{equation}
For a linear 1D arrangement of the four vortices one similarly obtains 
$g_{\_\_}=-V_0F_{1234}$.

To estimate $F_{1234}$ we must adopt some specific form for the radial part $f_0(\br)$ of the Majorana wavefunction, which in turn depends on the order parameter profile $\Delta(\br)$ near the vortex through Eq.\ (\ref{abel99}).
In the vicinity of a singly quantized vortex the latter is well approximated by \cite{tinkham1}
\begin{equation}\label{int6}
\Delta(\br)=\Delta_0\tanh{(r/\xi)},
\end{equation}
with $\Delta_0$ the asymptotic gap value far from the vortex. To facilitate analytical treatment we further expand Eq.\ (\ref{int6}) at small distances as
$\Delta(\br)\approx\Delta_0(r/\xi)$, which then leads to a normalized wavefunction
\begin{equation}\label{int8}
f_0(\br)\simeq{1\over \pi\xi}e^{-r^2/2\pi\xi^2},
\end{equation}
where we employed the BCS definition of the coherence length $\xi=v/\pi\Delta_0$. The above approximation is valid for $r\lesssim\xi$; for larger radii $f_0(\br)$ crosses over to a simple exponential dependence $\sim e^{-r/\xi}$. The advantage of the approximate form (\ref{int8}) is that the overlap integral in Eq.\ (\ref{int5}) is Gaussian and can be easily evaluated. For four vortices forming a square with a side of length $d$ one obtains
\begin{equation}\label{int9}
F_{1234}={1\over 2\pi^2\xi^2}e^{-d^2/\pi\xi^2}.
\end{equation}
For $d\gtrsim\xi$ one again expects a crossover to a simple exponential behavior $\sim e^{-d/\xi}$.

To complete the estimate we need the characteristic value of $V_0$. Assuming screened Coulomb interaction between electrons of the form $V_{\rm TF}(\br)=(e^2/r)e^{-r/\lambda}$ with $\lambda$ the Thomas-Fermi screening length, we find $V_0=\int d^2r V_{\rm TF}(\br)=2\pi e^2\lambda$. Putting everything together we thus arrive at an estimate
\begin{equation}\label{int10a}
g^{}_\square\simeq-{2e^2\lambda\over \pi\xi^2}e^{-d^2/\pi\xi^2}.
\end{equation}
A more transparent expression arises if we introduce the Bohr radius $a_0=\hbar^2/me^2\simeq 0.52\times 10^{-10}$m and the associated energy scale $\epsilon_0=e^2/2a_0\simeq 13.6$eV,
\begin{equation}\label{int10b}
g^{}_\square\simeq-\epsilon_0{4\over\pi}{a_0\lambda\over\xi^2}e^{-d^2/\pi\xi^2}.
\end{equation}
To estimate the typical interaction strength we take  the
experimentally measured  \cite{xu1}   coherence length $\xi\simeq 29$
nm. The value of the screening length $\lambda$ in this system is not
known but we note that it should be significantly longer than the screening  inside  a typical
metal (or a superconductor) because  in the setup of Ref.\ \cite{xu1}
the surface layer of the STI is separated from the SC substrate by the
insulating bulk of the STI crystal with thickness $h\simeq 3-10$
nm. The STI surface state itself should not screen efficiently because of
its low density of states.  A simple exercise in elementary
electrostatics shows that the screening length in this situation 
is then bound from below by distance $h$. This can be seen, for instance, by noting that the screening field can be attributed to  
the relevant image charge placed distance $h$ below the SC
surface. We can thus use distance $h$ as a rough estimate for the screening
length $\lambda\simeq 10$ nm to obtain an estimate for the interaction
strength $g^{}_\square\simeq (10.6 {\rm meV})\times e^{-d^2/\pi\xi^2}$. An even stronger interaction could be achieved in a material with a shorter coherence length or longer Thomas-Fermi screening length $\lambda$. Because of the exponential dependence on the intervortex distance $d$, the interaction effects will be most pronounced when $d$ does not exceed $\xi$ by a wide margin. For instance when $d=2\xi$ we obtain a respectable $g^{}_\square\approx 3$meV interaction scale. 

The interaction strength is to be compared with the direct hopping amplitude, which under the same assumptions as above becomes
\begin{equation}\label{int10c}
t_{12}\simeq\mu e^{-d^2/4\pi\xi^2}.
\end{equation}
Strong correlation regime obtains when $\mu$ is tuned such that $|t_{12}|\ll |g^{}_\square|$. In a typical experiment $\mu$ is controled by a combination of chemical doping and electrostatic gating. The latter is a continuous process in
which, presumably, the average $\mu$ can be tuned  as close to zero as desired. From this perspective, achieving the interaction dominated
regime should not present a significan problem, except of course that one must also ensure that the interaction effects are not obscured by disorder. We further discuss disorder effects in Sec.\ IV.C.

\section{Lattice models with interacting Majorana fermions}

We now proceed to study specific interacting models in one and two
spatial dimensions. We focus on lattice geometries whose building blocks are either 1D line segments or 
square plaquettes because they most naturally accommodate the four-fermion
interaction terms (\ref{int1}). We begin with 1D structures which can
be physically realized by inducing SC order in a narrow strip on the
surface of an STI and then applying magnetic field of appropriate
strength perpendicular to the surface. In 2D we focus on vortex
lattices with square symmetry. We note that although in most
conventional superconductors natural vortex lattices are triangular
\cite{tinkham1}, there exist materials with a strong four-fold
anisotropy in which square vortex lattices have been experimentally
observed  \cite{riseman1,gilardi1,curran1,yazdani1}. To engineer more
complex vortex structures one could also employ various techniques
that generate vortex pinning \cite{daldini1,baert1,harada1}. This
involves, essentially, perturbing the superconductor in a controlled
fashion on the nanoscale to create a pattern of regions with locally
suppressed SC order parameter $\Delta(\br)$. Such regions then attract
and  pin vortex cores due to the lower condensation energy. With sufficiently strong pinning one can, in principle, create almost arbitrary arrangement of vortices, including systems with e.g.\ multiply quantized vortices which are otherwise energetically unstable.

\subsection{One-dimensional lattice models}

One may expect on symmetry grounds that one-dimensional vortex lattice structures will arise when a strip of a superconducting thin-film material is deposited on the STI surface and subjected to a perpendicular magnetic field. 
Theoretical calculations within the Ginzburg-Landau theory indeed predict a single line of vortices forming along the long axis of the strip at low fields and more complicated structures with multiple lines at higher fields \cite{teniers1}. Some of these predictions have been confirmed experimentally \cite{morelle1}. Importantly, these calculations also indicate that the intervortex distance in such configurations is typically much smaller than the distance between vortices and the strip edge. This means that the interactions between Majorana zero modes bound to vortices will dominate over any residual interactions with low-energy Dirac fermions present in the ungapped surface of the STI (note also that the density of states of the latter vanishes when $\mu\approx 0$).

\subsubsection{Linear chain}

A simple {\em linear chain} depicted in Fig. \ref{fig2}a is described
by an interacting Hamiltonian of the form
\begin{equation}\label{lin1}
{\cal H}_{\rm int} =g_1\sum_{j} \alpha_j\beta_j\alpha_{j+1}\beta_{j+1}
+g_2\sum_{j} \beta_j\alpha_{j+1}\beta_{j+1}\alpha_{j+2}.
\end{equation}
Here $\alpha_j$ and $\beta_j$ denote two Majoranas in the two site
unit cell $j$. In a uniform chain $g_1=g_2$ but we consider here a more general case
of dimerized bond lengths leading to alternating couplings $g_1$ and $g_2$. The Hamiltonian (\ref{lin1}) can be brought to a more
familiar form by performing a Wigner-Jordan transformation suitable for Majorana fermions \cite{fidkowski1} to spin
variables ${\bm \sigma}_j$,
\begin{equation}\label{wj}
\alpha_j=\left(\prod_{k=1}^{j-1}\sigma^x_k\right)\sigma^z_j, \ \ \ 
\beta_j=i \left(\prod_{k=1}^{j-1}\sigma^x_k\right)\sigma^z_j\sigma^x_j.
\end{equation}
One obtains
\begin{equation}\label{lin2}
{\cal H}_{\rm int} =-g_1\sum_{j} \sigma^x_j\sigma^x_{j+1}
-g_2\sum_{j} \sigma^z_j\sigma^z_{j+2},
\end{equation}
an interesting variant of the XY model, with nearest neighbor spin
interactions along $x$ and next nearest interactions along $z$. This
is an example of a spin model that would not naturally arise in a
system where fundamental degrees of freedom are electron spins. Yet it
emerges here from a very simple and natural structure composed of interacting
Majorana fermions.

Adding direct hopping terms (assuming again a dimerized lattice)
described by  
\begin{equation}\label{lin3}
{\cal H}_{\rm kin} =it_1\sum_{j} \alpha_j\beta_j
+it_2\sum_{j} \beta_j\alpha_{j+1}
\end{equation}
gives, in the spin representation,
\begin{equation}\label{lin4}
{\cal H}_{\rm kin} =-t_1\sum_{j} \sigma^x_j
-t_2\sum_{j} \sigma^z_j\sigma^z_{j+1}.
\end{equation}
The full Hamiltonian ${\cal H}={\cal H}_{\rm int} + {\cal H}_{\rm
  kin}$ is not exactly solvable for a general set of parameters but has several special points in the
parameter space where exact solutions are known. These include an anisotropic XY model for
$g_2=t_1=0$ and a transverse field Ising model when $g_1=t_2=0$ or
when $g_1=g_2=0$. A detailed exploration of the phase diagram of this
model is beyond the scope of this study and we leave it to future
work. It is clear, however, that the model exhibits a rich phase
diagram with gapped and gapless phases, some of which are
topologically non-trivial and carry unpaired Majorana zero modes at
the edges.

\subsubsection{Two-leg ladder}

Next we consider a {\em two-leg ladder}  shown in Fig. \ref{fig2}b. The
interacting Hamiltonian is given by the first term in Eq.\
(\ref{lin1}).  This model
is exactly solvable for an arbitrary interaction strength $g$ on the
square plaquette. To see this note that each four-fermion term
commutes with the Hamiltonian and is therefore a constant of
motion. Furthermore, adding hopping $t$ along the rung does not spoil
the model's integrability, although hopping $t'$ along the legs
does. After the WJ transformation (\ref{wj}) the Hamiltonian can be
written as ${\cal H}={\cal H}_0 + {\cal H}'$ with 
\begin{eqnarray}\label{lad1}
{\cal H}_0&=&-g\sum_j\sigma^x_j\sigma^x_{j+1}-t \sum_j\sigma^x_j, \\
{\cal H}'&=&-t'\sum_j\left(\sigma^y_j\sigma^z_{j+1}-\sigma^z_j\sigma^y_{j+1}\right).
\end{eqnarray}
The signs of $t$ and $t'$ terms here reflect the Z$_2$ gauge factors indicated in Eq.\ (\ref{kin2}).
At the neutrality point ($t=t'=0$) and assuming $g>0$ the ground state
is a doubly degenerate ferromagnet. In the fermion language this
corresponds to complex fermions $c_j={1\over 2}(\alpha_j+i\beta_j)$ on each rung
either all occupied or all empty. Turning on $t\neq 0$ removes the
two-fold degeneracy. Since this is a gapped state one expects it to remain
stable against the perturbation ${\cal H}'$ as long as $t'$ remains weak.

\subsubsection{Diamond chain}

As the final 1D example we consider a {\em diamond chain} depicted in
Fig. \ref{fig2}c. The interacting Hamiltonian is 
\begin{equation}\label{dia1}
{\cal H}_{\rm int} =g_1\sum_{j\  {\rm odd} } \gamma_{j} \alpha_j\beta_j\gamma_{j+1}
+g_2\sum_{j\  {\rm odd}} \gamma_{j+1}\alpha_{j+1}\beta_{j+1}\gamma_{j+2},
\end{equation}
where once again we allow for the possibility of dimerization. We
observe that products $\alpha_j\beta_j$ commute with ${\cal H}_{\rm
  int}$ 
and with one another. They can thus be replaced by classical
variables $is_j=\pm i$.  The Hamiltonian becomes
\begin{equation}\label{dia2}
{\cal H}_{\rm int} =i\sum_{j\  {\rm odd} }\left(g_1s_j \gamma_{j} \gamma_{j+1}
+g_2s_{j+1}\gamma_{j+1}\gamma_{j+2}\right),
\end{equation}
describing a simple 1D chain with hoppings $g_1$ and $g_2$ between
nearest neighbor sites. Because there are no closed loops in such a
linear chain we can adopt a gauge in which $s_j=1$ for all $j$.  The
Hamiltonian (\ref{dia2}) then coincides with the Kitaev chain model
\cite{kitaev1}. Accordingly, its spectrum is gapped whenever $g_1\neq
g_2$. For an open ended chain with sites labeled $j=1 \dots 2N$ the
phase with $g_1<g_2$ is topological and has unpaired Majorana zero
modes bound to its two ends while $g_1>g_2$ corresponds to the trivial
phase. $g_1=g_2$ marks the critical point separating the two phases.
Adding ${\cal H}_{\rm kin}$ to the interacting Hamiltonian
(\ref{dia1}) spoils its integrability but again we may expect the
gapped phases to be robust against small detuning from the neutrality
point.

\subsection{Two-dimensional lattice models}

We now turn to 2D lattice geometries. A simple square lattice depicted in Fig.\ \ref{fig2}d is not exactly solvable and we shall discuss its phase diagram below. We consider first a modified square lattice shown in Fig.\ \ref{fig2}e which represents a somewhat artificial but exactly solvable 2D geometry with strong interactions. It is obtained by populating one sublattice with doubly quantized vortices each containing two exact Majorana zero modes $\alpha_{\bj}$, $\beta_{\bj}$. The dominant interaction terms in this arrangement are of the form
\begin{equation}\label{sq1}
{\cal H}_{\rm int} =g\sum_{\bj,{\bm \nu}}\alpha_{\bj}\beta_{\bj}\alpha_{\bj+{\bm \nu}}\beta_{\bj+{\bm \nu}} 
+ g'\sum_{\bj,{\bm \delta}_1,{\bm \delta}_2}\alpha_{\bj}\beta_{\bj}
\gamma_{\bj+{\bm \delta}_1}\gamma_{\bj+{\bm \delta}_2}
\end{equation}
where 
${\bm \delta}=\pm\hat{x},\pm\hat{y}$ are the nearest neighbor vectors while ${\bm \nu}$ second neighbor vectors on the square lattice. The model is solvable because once again products  $\alpha_{\bj}\beta_{\bj}$ commute with ${\cal H}_{\rm  int}$ (and with one another) and can thus be replaced by classical variables $is_{\bj}=\pm i$. The resulting Hamiltonian is bilinear in the $\gamma$ operators residing on the single vortex sites and can be analyzed in a straightforward fashion. Depending on the relative sign and amplitude of the couplings $g$ and $g'$ various phases are possible, including a gapless metallic phase when $g\gg g'>0$ and, interestingly, dispersionless flat band at zero energy when $g<0$ and $|g|\gg|g'|$. A detailed discussion of this model is given in Appendix \ref{single double}.

\begin{figure*}[t]
\includegraphics[width = 16.0cm]{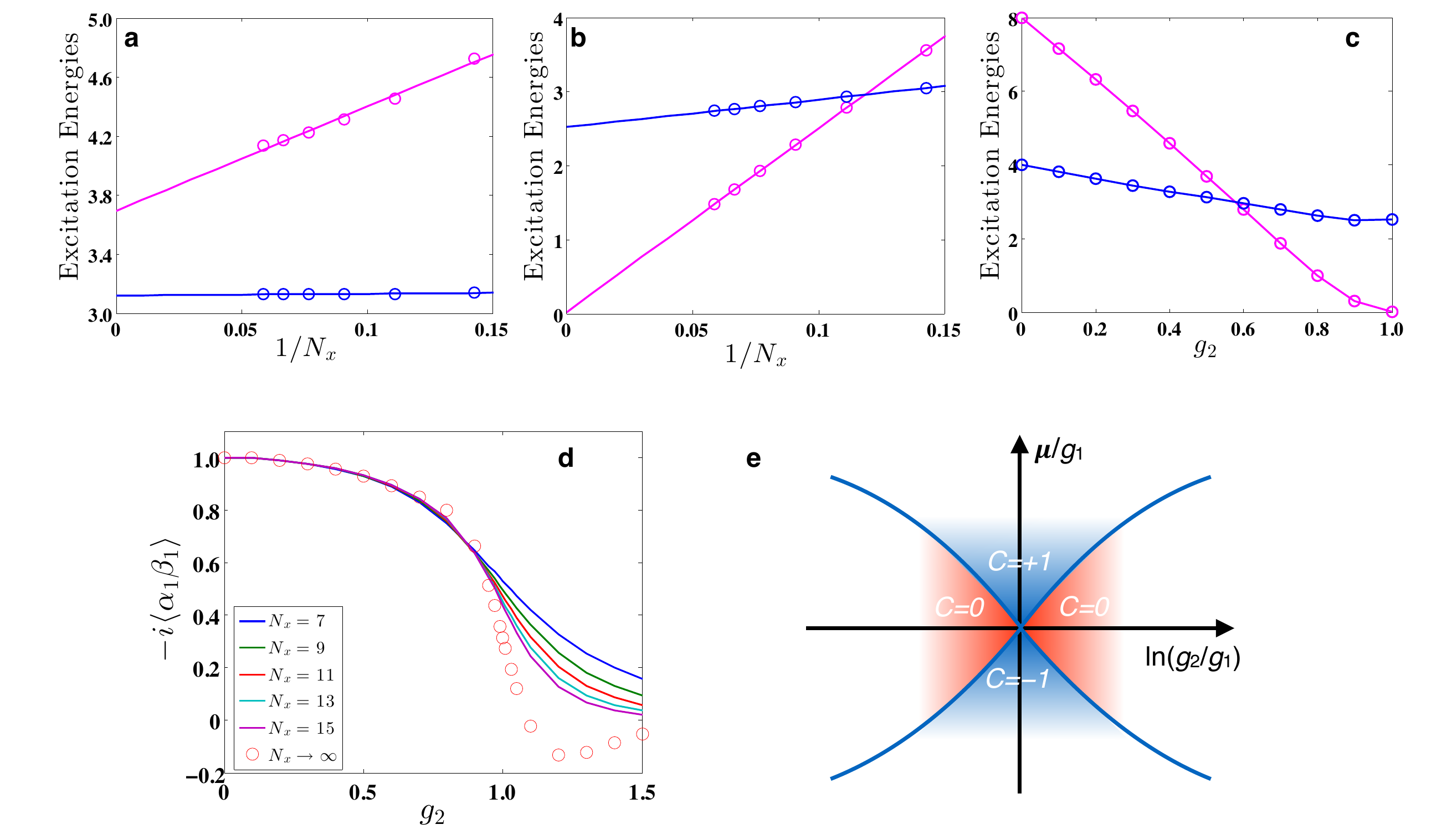}
\caption{ Interacting system of Majorana fermions on the simple square lattice. Panels {\bf a-c} show the finite size scaling analysis of the many-body excitation energies of the system obtained by exact numerical diagonalization. Details of the numerical procedure are described in Appendix \ref{single ED}. Energies of the two lowest excited states are plotted as a function of $1/N_x$ for  $g_2=0.5$ in {\bf a} and $g_2=1.0$ in {\bf b}. The excitation energies extrapolated to $N_x\to \infty$ are displayed in panel {\bf c} as a function of $g_2$. This plot shows that the gap closes at $g_2=g_1$ indicating a phase transition. We note that the first excited state here exhibits degeneracy that grows with a system size.
Panel {\bf d} shows the order parameter $\Delta_1=-i\langle\alpha_1\beta_1\rangle$ as a function of $g_2$  for various system sizes. The infinite system extrapolation is obtained by assuming $\Delta_1\simeq c_0+c_1/N_x+c_2/N_x^2$. The order parameter goes to zero continuously at $g_2=g_1$ supporting the notion of the continuous phase transition. In panels {\bf a-c} $g_1=1.0$ is held constant. {\bf e} The schematic phase diagram for the simple square lattice system.    
}\label{fig3}
\end{figure*}
A simple square lattice model depicted in Fig.\ \ref{fig2}d cannot be reduced to a non-interacting problem and we study it by a combination of approximate analytical techniques and by exact numerical diagonalization on small clusters. To facilitate the discussion we consider a dimerized situation with couplings $g_1$ and $g_2$ on alternating rows of plaquettes, described by  
\begin{equation}\label{sq2}
{\cal H}_{\rm int} =g_1\sum_{\bj}\alpha_{\bj}\beta_{\bj}\alpha_{\bj+{\bm x}}\beta_{\bj+{\bm x}} 
+ g_2\sum_{\bj}\beta_{\bj}\alpha_{\bj-{\bm y}}\beta_{\bj+ {\bm x} }\alpha_{\bj+ {\bm x}-{\bm y}}
\end{equation}
In the limit $g_2=0$ the system breaks up into a collection of two-leg ladders already discussed above. Assuming $g_1,g_2\geq 0$ the exact ground state is a direct product of the ground states of the individual ladders. In the language of Ising spins defined in Eq.\ (\ref{wj}) these are doubly degenerate 1D ferromagnets. The ground state thus exhibits a $2^{N_y}$-fold degeneracy, where $N_y$ is the number of unit cells in the $y$ direction. The spectrum of excitations is gapped and the lowest excited state at energy $2g_1$ has one of the spins reversed. Inclusion of nonzero $g_2$ can be seen to suppress the ferromagnetic order in the individual ladders by promoting excitations. A reasonable conjecture is that the gapped phase persist all the way to the isotropic point $g_2=g_1$ which marks a quantum phase transition to another gapped state that is adiabatically connected to a set of independent ladders that occur at $g_1=0$.    

We have performed a standard mean-field (MF) analysis by decoupling ${\cal H}_{\rm int}$ in all possible channels involving Majorana bilinears on nearest and next nearest neighbor bonds. At $g_2=0$ this procedure yields the exact ground state with $\Delta_1=g_1\langle i\alpha_{\bj}\beta_{\bj}\rangle=\pm g_1$ and all other order parameters zero. The two possible signs correspond to two degenerate ferromagnetic ground states on each ladder. Interestingly, this solution persists as the mean-field ground state for all values of $g_2<g_1$. At $g_2=g_1$ the MF theory predicts a strong first order transition to a state characterized by non-vanishing order parameter $\Delta_2=g_2\langle i\beta_{\bj}\alpha_{\bj-{\bm y}}\rangle=\pm g_2$ which then persists all the way to $g_1=0$ where it becomes the exact ground state of ${\cal H}_{\rm int}$. To ascertain the accuracy of the MF solution we carried out exact numerical diagonalizations (ED) of ${\cal H}_{\rm int}$ for a system containing $N_x\times 4$ lattice sites with $N_x$ up to 19 (see Appendix \ref{single ED}). Some representative results are displayed in Fig.\ \ref{fig3}. These indicate that MF treatment provides a reasonable approximation for $g_2/g_1\ll 1$ but breaks down when the two couplings are comparable. Specifically, ED indicates a continuous phase transition at $g_2=g_1$ with the gap closing smoothly at that point.

We expect the gapped phases of the 2D model to remain robust against small detuning from the neutrality point. However, at the criticality, such detuning is likely to drive the system into another phase, adiabatically connected to the noninteracting system of Majorana fermions described by Hamiltonian (\ref{kin1}). Our conjectured phase diagram describing this situation is displayed in Fig. \ref{fig3}e. The gapped phases in the interaction dominated regime are separated from the hopping dominated phases by topological phase transitions. This can be seen by analyzing the noninteracting Hamiltonian (\ref{kin1}). It describes spinless fermions with charge conjugation symmetry. Since the time reversal symmetry is absent the system is in topological class D which has integer classification in $d=2$. Assuming that $t_{ij}$ is dominated by first and second neighbor hoppings $t$ and $t'$ the system is gapped and one can easily calculate the corresponding Chern number $C={\rm sgn}(t^2t')=\pm 1$. Recalling furthermore that for small chemical potential $t,t'\propto\mu$, as shown in Sec. II, we obtain  $C={\rm sgn}(\mu)$, leading to the phase diagram illustrated in Fig. \ref{fig3}e. The interaction dominated phases by contrast are adiabatically connected to systems of decoupled two leg ladders and are thus topologically trivial with $C=0$. 

\subsection{Physical feasibility and proposed experimental observations}
Models discussed in this Section can be engineered in a laboratory
provided that several conditions are met. The key requirement is the
ability to tune the chemical potential $\mu$ of the STI surface state to the
close vicinity of the neutrality point. Although the most common STIs in the
Bi$_2$Se$_3$ family do not naturally grow in this regime, neutrality
point can be reached in these via chemical doping and by electrostatic
gating in
the thin film or flake geometry. Remarkably, tantalizing evidence for
intrinsic  surface superconductivity with $T_c\simeq 9$K and
$\Delta_0\simeq 5$ meV has recently
been reported \cite{elbaum1}  in topological insulator Sb$_2$Te$_3$
whose growth chemistry has been tuned to achieve neutrality. Although
the mechanism behind the emergence of superconducting order in this
material  is presently not known, if confirmed this system could
form an ideal platform for the exploration of the lattice models with
interacting Majorana  fermions.  In other, more recently discovered  STI materials, such
as the ternary  Bi$_2$Te$_2$Se,  the $\mu\approx 0$ condition naturally obtains in a
stoichiometric crystal \cite{ando1,cava1}.  Quaternary compounds
Bi$_{2-x}$Sb$_x$Te$_{3-y}$Se$_y$ can in turn be robustly tuned into
their neutrality point \cite{ando2}.  

The samples must also be sufficiently clean so that the interaction effects
are not obscured by disorder. The situation here resembles
fractional quantum Hall systems where the sample quality is of paramount
importance.  Disorder that breaks the chiral symmetry of the Fu-Kane
model, such as the fluctuating scalar potential, will generate random
Majorana hopping between the adjacent vortices. These must be
negligible compared to the interaction scale $g$ that we estimated to
be of the order of several meV. Disorder that does not break the symmetry,
such as irregularities in the vortex positions or fluctuations in the SC
pairing amplitude, will not generate hopping
terms but will introduce a random component $\delta g$ in the
interaction strengths. Understanding the effect of disorder in a
strongly interacting system is a difficult problem, one that lies
beyond the scope of this study. By thinking about those interacting
models that are exactly solvable (such as the two leg ladder and the
diamond chain) we may conclude that weak disorder  $|\delta g| \ll
|g|$ will have negligible effect on the gapped phases but could affect
the nature of the critical points in some cases. In models that are
not integrable disorder could lead to more interesting phenomena such
as the many-body localization. This, obviously, is a potentially
interesting topic for future studies.

The most obvious experimental tool to probe the interacting
systems we described in this study is scanning tunneling microscopy
(STM). This technique is uniquely suited to image vortex lattices at
the nanoscale \cite{yazdani1} as well as to detect bound states
present in the vortex cores \cite{gygi1,magio1}. A first step towards observing
the complex phenomena associated with interactions will be to resolve
a single Majorana zero mode in the vortex core of the Fu-Kane model and
its splitting as a result of hybridization with another zero mode
localized in a nearby vortex. We note that once a suitable sample with
$\mu\approx 0$ has been fabricated this should be a relatively easy task because in
this limit Fu-Kane model predicts a single vortex core state at zero
energy separated  from all other core states by a gap whose amplitude
is close to the full SC gap $\Delta_0$ \cite{cook1,cook2}. With the SC gap
of the order of meV, as seen in Ref.\ \cite{xu1},  a state of the art STM should have
no problem clearly resolving the zero modes and their splitting due to
hybridization or interaction effects.

Once the zero modes are detected the next step will consist of
establishing the effect of interactions in small clusters of
vortices. This again, should be relatively
straightforward. Interaction effects are easy to distinguish from
simple hybridization because they require four or more vortices to
occur. Thus, a smoking gun test for the interaction effect is to probe the
zero mode splitting  in a group of 2, 3 and 4
vortices. Hybridization, if present, will split the zero modes in all
cases while interaction will only cause splitting in the last
case. When the interaction effect is confirmed in such small clusters
then one can move onto larger lattices which will, for correct
geometries, show interesting collective phenomena.

We have discussed in this Section some specific examples of vortex lattice geometries
that lead to simple interacting models with Majorana fermions. Even
these basic structures display
interesting behaviors.  The actual experimental vortex lattice
geometries will depend on the details of the physical samples and we shall not
attempt  here to specify the precise conditions for the formation of a
given structure. Instead, we note that since STM can be used to map
out both the lattice structure and the electronic state of vortices,
 theory will work best in conjunction with experiment to unveil the
 physics of strong interactions in these systems.

\section{Outlook}

When the chemical potential is tuned to coincide with the Dirac point
in the superconducting surface of a strong topological insulator
Majorana fermions bound to the vortex cores show a completely flat
band, protected by the chiral symmetry. In this regime the nature of
the ground state is determined by interactions between the Majorana
zero modes and the system must be regarded as inherently strongly
correlated. We gave examples of lattice geometries in one and two
dimensions for which the ground state of the strongly interacting
system can be found exactly. In other cases, such as the simple 1D
Majorana chain, exact solution of the interacting problem is unknown
but the Hamiltonian maps onto an interesting spin problem which can be
studied by standard techniques such as the density matrix
renormalization group (DMRG). Although well understood theoretically
spin models in 1D often face significant hurdles when it comes to
their experimental realizations. For instance the fine details of quantum
criticality in the transverse field Ising model -- perhaps the most
widely studied 1D spin model -- have been only recently mapped out
experimentally \cite{coldea1}. Our construction may thus enable new
experimental realizations of these well studied models. In addition,
it may help realize spin models that do not naturally occur in systems
whose fundamental degrees of freedom are spins, as in the case of the
interacting 1D Majorana chain.

Interesting phenomena occur  also in two-dimensional systems.
  The simple square lattice  shows an intriguing phase diagram with both topological and trivial gapped phases as well as a quantum phase transition that cannot be described by  mean field theory. Further interesting phases in 2D may arise in lattices with triangular symmetry which we have not considered in this study.

Physical realizations of interacting systems with Majorana fermions in some respects similar to ours have been previously discussed in the context of semiconductor quantum wire networks \cite{hassler1,terhal1,kells1}. The existence of Majorana fermions in the individual quantum wires has been established by recent ground breaking experiments \cite{mourik1,das1,deng1,rohkinson1,finck1,churchill1,Ramon}. However, assembling these into large arrays with uniform properties and tunable interaction and hopping parameters appears to be a much more difficult challenge, one that will likely require new experimental methodologies.  By contrast, scaling the systems of few vortices with Majorana zero modes, such as those observed in Bi$_2$Te$_3$/NbSe$_2$ heterostructures \cite{xu1}, to large lattices required in our proposal seems to be rather straightforward. The key issue that must be surmounted to achieve the strong correlation regime here is the ability to tune the system  to its global neutrality point. In addition, local fluctuations of the chemical potential must remain sufficiently small as to render disorder effects negligible compared to the interaction energy scale. We estimated in Sec. III that the characteristic interaction energy in Bi$_2$Te$_3$/NbSe$_2$ heterostructures is  $\sim 10$ meV. We emphasize that only disorder strength  averaged over distances comparable to intervortex spacing $d$ (of the order of $10-100$ nm) must be small compared to the interaction energy, which should be achievable in clean STI samples.   

The ultimate goal of these constructions is to find novel phases that cannot be adiabatically deformed into phases of weakly interacting fermions or interesting phase transitions that do not have a free particle description.   
That such phases or transitions can indeed occur in these systems could  be anticipated because Majorana interactions of the form Eq. (\ref{int1}) play a pivotal role in the construction of various ``interaction enabled'' topological phases introduced in the seminal work by Fidkowski and Kitaev \cite{fidkowski1}. Our work indicates how such interactions can be generated and controlled in a system that is now physically accessible thanks to the recent experimental breakthroughs \cite{koren1,elbaum1,sacepe1,qu1,williams1,cho1,xu0,xu1}. We note that recently a specific model has been formulated by Lapa, Teo and Hughes \cite{lapa1} that produces an interaction enabled topological crystalline phase (which has no analog in a weakly interacting system) and also employs Majorana interaction of the type discussed in this work as the key component. One can show that such a phase can be in fact constructed from the ingredients introduced in this study \cite{chiu2}.

\section{Acknowledgment}

The authors thank I. Affleck, J. Alicea, T. Liu, A. Rahmani, G. Refael, K. Shtengel and X. Zhu for useful discussions.
The authors are indebted to  NSERC, CIfAR and Max Planck - UBC Centre for Quantum Materials for support. M.F.\ acknowledges The Aspen Center for Physics and IQMI at Caltech for hospitality during various stages of this project.

\appendix

\section{Phase factors, branch cuts and the Z$_2$ gauge structure}\label{phases}
In this Appendix we outline the computation of the relevant phase factors that enter the overlap integrals for Majorana zero modes in Eq.\ (\ref{abel112}) and the interaction amplitudes (\ref{abel16}). We also explain how the Z$_2$ gauge factors that appear in the Majorana tight  binding model arise from branch cuts present in the vortex lattice.

\begin{figure}[t]
\includegraphics[width = 8.0cm]{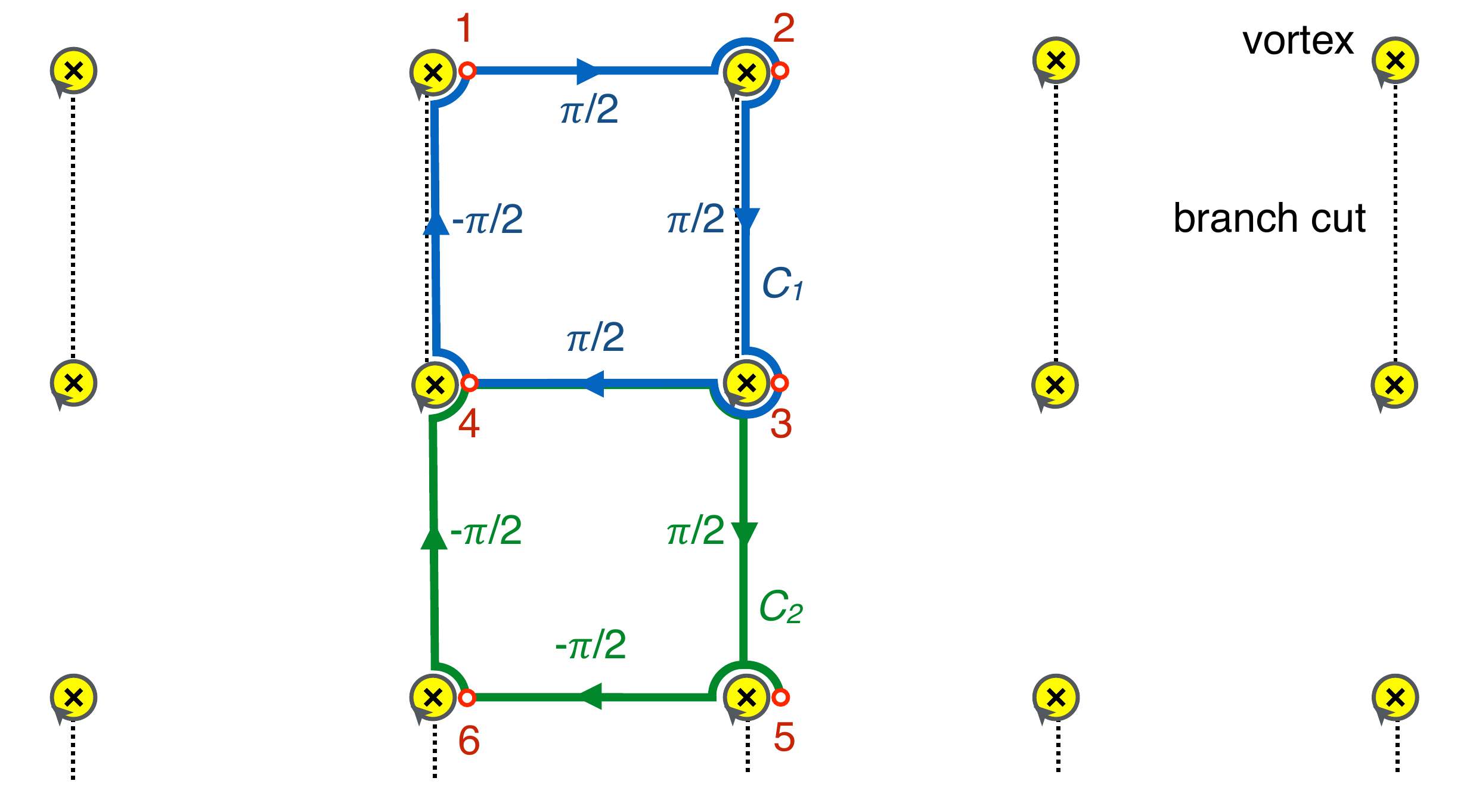}
\caption{Phase factors and branch cuts in a square vortex lattice. Oriented solid lines indicate integration paths between the reference points located just to the right of each each vortex center. Dashed lines represent a specific choice of the branch cuts discussed in the text.  
}\label{fig100}
\end{figure}

Although the method outlined here is applicable to an arbitrary arrangement of vortices, we focus, for the sake of concreteness, on a periodic vortex lattice such as the one depicted in Fig.\ \ref{fig100}. Following \cite{biswas1} we define the phase $\theta_j$ that enters the definition of the Majorana wavefunction (\ref{abel9}) at a point $\br_j$ immediately to the right of the given vortex center, to avoid the phase singularity. The overlap integral between the two vortices at $\br_i$ and $\br_j$ is then given, according to Eq.\ (\ref{abel12}), as $t_{ij}=\mu F_{ij}\sin{\omega_{ij}}$ with 
\begin{equation}\label{a1}
\omega_{ij} =\int_{\br_i}^{\br_j}\left({1\over 2}\nabla\theta-{e\over
    \hbar c}\bA\right)\cdot d{\bf l},
\end{equation}
where we have restored $\hbar$. The integrand in Eq.\ (\ref{a1}), which we henceforth call $\bOmega$, is closely related to the superfluid velocity \cite{tinkham1}
\begin{equation}\label{a2}
\bv_s={\hbar\over m^*}\left(\nabla\theta-{e^*\over\hbar c}\bA\right)=
{2\hbar\over m^*}\bOmega.
\end{equation}
Here $e^*=2e$ and $m^*$ are, respectively, the effective charge and mass of the Cooper pair. The superfluid velocity distribution in the vortex lattice can be calculated in a straightforward way \cite{tinkham1}  which we review below for completeness. It is related to the supercurrent $\bj_s=e^*n_s\bv_s$ where $n_s$ represents the superfluid density. 

The calculation proceeds by taking the curl of $\bj_s$, 
\begin{equation}\label{a3}
\nabla\times\bj_s=n_s{e^*\hbar\over m^*}\left(\nabla\times\nabla\theta-{e^*\over\hbar c}\bB\right),
\end{equation}
and noting that 
\begin{equation}\label{a4}
\nabla\times\nabla\theta=2\pi\hat{z}\sum_j\delta(\br-\br_j),
\end{equation}
where $\br_j$ are the vortex positions and we are assuming that the SC interface lies in the $x-y$ plane. We now use the Amp\`{e}re's law $\nabla\times\bB=(4\pi/c)\bj_s$ to eliminate the current from Eq.\ (\ref{a3}). We thus find the London equation for $\bB$ in the vortex lattice,
\begin{equation}\label{a5}
\bB-\lambda_L^2\nabla^2\bB={1\over 2}\Phi_0\hat{z}\sum_j\delta(\br-\br_j),
\end{equation}
where $\lambda_L^2=mc^2/4\pi{e^*}^2 n_s$ is the London penetration depth and $\Phi_0=hc/e$ the flux quantum. For a periodic lattice the equation can be solved by Fourier transforming, 
\begin{equation}\label{a6}
\bB(\br)={1\over 2}\Phi_0\hat{z}\sum_\bG{e^{i\bG\cdot\br}\over 1+\lambda_L^2G^2},
\end{equation}
where the sum extends over all reciprocal vectors $\bG$ of the vortex lattice. From the knowledge of $\bB$ one can reconstruct the supercurrent via Eq.\ (\ref{a3}) and from it $\bv_s$. Finally,
\begin{equation}\label{a7}
\bOmega(\br)=\pi\sum_\bG{i\bG\times\hat{z}\over \lambda_L^{-2}+G^2}e^{i\bG\cdot\br}.
\end{equation}
The gauge invariant phase factors $\omega_{ij}$ can now be determined by a straightforward integration of $\bOmega(\br)$ indicated in Eq.\ (\ref{a1}).

The above method works for any vortex lattice but 
in cases with high symmetry, such as the square lattice, the phase factors can be deduced without performing a detailed calculation. Consider the lattice depicted in Fig.\ \ref{fig100}. The integration paths between points $\br_j$ have been chosen to consist of straight line segments and circular segments. The latter are needed to avoid the phase singularities located at each vortex center. In the following we think of these as having an arbitrarily small radius so that the contribution to the line integral along the cirgular segment comes exclusively from the adjacent singularity. Now consider the path $C_1$ indicated in Fig.\ \ref{fig100}. The corresponding line integral $\oint_{C_1}\bOmega\cdot d{\bf l}=\int(\nabla\times\bOmega)\cdot d\bS$ can be seen to equal to $\pi$; it encloses two vortices, each contributing flux $\pi$ and a half quantum of magnetic flux in the opposite direction contributing $-\pi$. We furthermore note that counting just the contribution of the  circular line segments around the vortices one gets the same answer $\pi$ for the total flux. We are thus led to a conclusion that the straight line segments do not contribute to $\omega_{ij}$. This same conclusion can be reached by similarly examining the path $C_2$ which contains total flux $-\pi$. Thus, in the square vortex lattice, we can determine the phase factors $\omega_{ij}$ by simply adding up the contributions from the circular segments around the individual vortices, which are given by their angular length divided by two. This leads to the phase factors $\omega_{ij}=\pm\pi/2$ indicated in the Figure, a result that can be confirmed by an explicit calculation using Eq.\ (\ref{a7}). 

The above arguments contain an important subtlety that has to do with branch cuts. Consider for instance the path indicated in Fig.\ \ref{fig100} between points 1 and 2. Had we chosen a path avoiding the vortex from below (instead of going above it) we would have found the phase to be $-\pi/2$. More generally, $\omega_{ij}$ changes to  $\omega_{ij}\pm\pi$, depending on which way we decide to avoid the singularity. The magnitude of $t_{ij}$ is independent of this choice but its sign depends on it because $\sin{(\omega_{ij}\pm\pi)}=-  \sin{\omega_{ij}}$. This is the origin of the Z$_2$ gauge structure in Eq.\ (\ref{kin1}). The latter is inherent to the tight binding models with Majorana fermions and arises here from the physics of branch cuts. In order to consistently determine the signs of $t_{ij}$, which become physically relevant when there exist closed loops in the model, one must define $\omega_{ij}$ in a globally unique fashion. This can be done by specifying branch cuts across which $\bOmega(\br)$ changes discontinuously. A branch cut emanates from each vortex core and can be chosen to terminate in another vortex core. An example of a specific choice of branch cuts is given in Fig.\ \ref{fig100}. Integration paths that do not intersect any branch cuts then furnish a globally consistent definition of the gauge invariant phase factors $\omega_{ij}$. Different choices of branch cuts correspond to different Z$_2$ gauges for Majorana fermions, but they leave the physical observables unchanged. The phase factors indicated in Fig.\ \ref{fig100} have been obtained in accord with this prescription. They define a periodic lattice  with two vortices per unit cell and are consistent with the Grosfeld-Stern rule Eq.\ (\ref{kin2}). The same phase factors are used for the computation of the interaction amplitudes in Sec.\ III.

\section{Exactly solvable 2D model}\label{single double}

The building block for the solvable Majorana model in 2D is a doubly quantized  vortex defined by Eq.\ (\ref{abel4}) with $(n=2)$. The solution for the Majorana wavefunction goes along similar lines as for the single vortex \cite{rossi1}.
We search for zero mode solutions of operator $D$ defined in Eq.\ (\ref{abel8}) with $n=2$ in the form
\begin{align}
\chi_m(\mathbf{r}) = \frac{1}{\sqrt{2}}\begin{pmatrix}
e^{i((1 - m)\varphi + \theta/2 - \pi/4)} u_m(r)\\
e^{-i(m\varphi + \theta/2 - \pi/4)} v_m(r)
\end{pmatrix}.
\end{align}
We substitute this into $D$ to obtain
\begin{align}
\left\{\begin{array}{c}
\Delta_0(r) u_m(r) + \left(\partial_r - \frac{m}{r}\right) v_m(r) = 0,\\
\Delta_0(r) v_m(r) + \left(\partial_r - \frac{1 - m}{r}\right) u_m(r) = 0.
\end{array}\right.
\end{align}
It is known \cite{rossi1} that these equations have normalizable real solutions for $m=0, 1$, for which it holds
\begin{align}
u_1 = v_0, \; v_1 = u_0.
\end{align}
This observation allows us to write the field operator of the zero modes
\begin{eqnarray}
\alpha(\br) &\propto&[e^{i (\varphi + \theta/2 - \pi/4)} c_{\br \downarrow} + e^{-i (\varphi + \theta/2 - \pi/4)}  c_{\br\downarrow}^\dag]u_0(r) \nonumber \\ &+& [e^{i (\theta/2 - \pi/4)} c_{\br \downarrow} + e^{-i (\theta/2 - \pi/4)}  c_{\br \downarrow} ^\dag]v_0(r), \\
\beta(\br) &\propto &i[e^{i (\varphi + \theta/2 - \pi/4)} c_{\br \downarrow} \nonumber - e^{-i (\varphi + \theta/2 - \pi/4)}  c_{\br\downarrow}^\dag]u_0(r) \nonumber \\ & -& i[e^{i (\theta/2 - \pi/4)} c_{\br \downarrow} - e^{-i (\theta/2 - \pi/4)} c_{\br \downarrow} ^\dag]v_0(r). \nonumber
\end{eqnarray}
It is easy to show that the density is then given by
\begin{align}
\rho_{\alpha \beta} \propto [u_0^2(r) - v_0^2(r)].
\end{align}
This expression depends only on the distance from the vortex core. It decays exponentially on distances longer than the coherence length $\xi$.

We are now interested in the dominant interactions between the Majoranas in such a model. For this we notice that the interaction is the largest  for the combinations $g\alpha_{\bj}\beta_{\bj}\alpha_{\bj+{\bm \nu}}\beta_{\bj+{\bm \nu}}$ and $g'\alpha_{\bj}\beta_{\bj}
\gamma_{\bj+{\bm \delta}_1}\gamma_{\bj+{\bm \delta}_2}$ depending on how strong the screening of the Coulomb interactions is. The corresponding interaction strengths are proportional to $\exp\left[-|\mathbf{R}_{\bj + {\bm \nu}} - \mathbf{R}_\bj|/R_c\right]$, where $R_c$ is the Coulomb screening length, and $\exp\left[-(|\mathbf{R}_{\bj } - \mathbf{R}_{\bj+{\bm \delta}_1}| +  |\mathbf{R}_{\bj } - \mathbf{R}_{\bj+{\bm \delta}_2}|)/\xi\right]$.

\begin{figure}[t]
\includegraphics[width = 8.0cm]{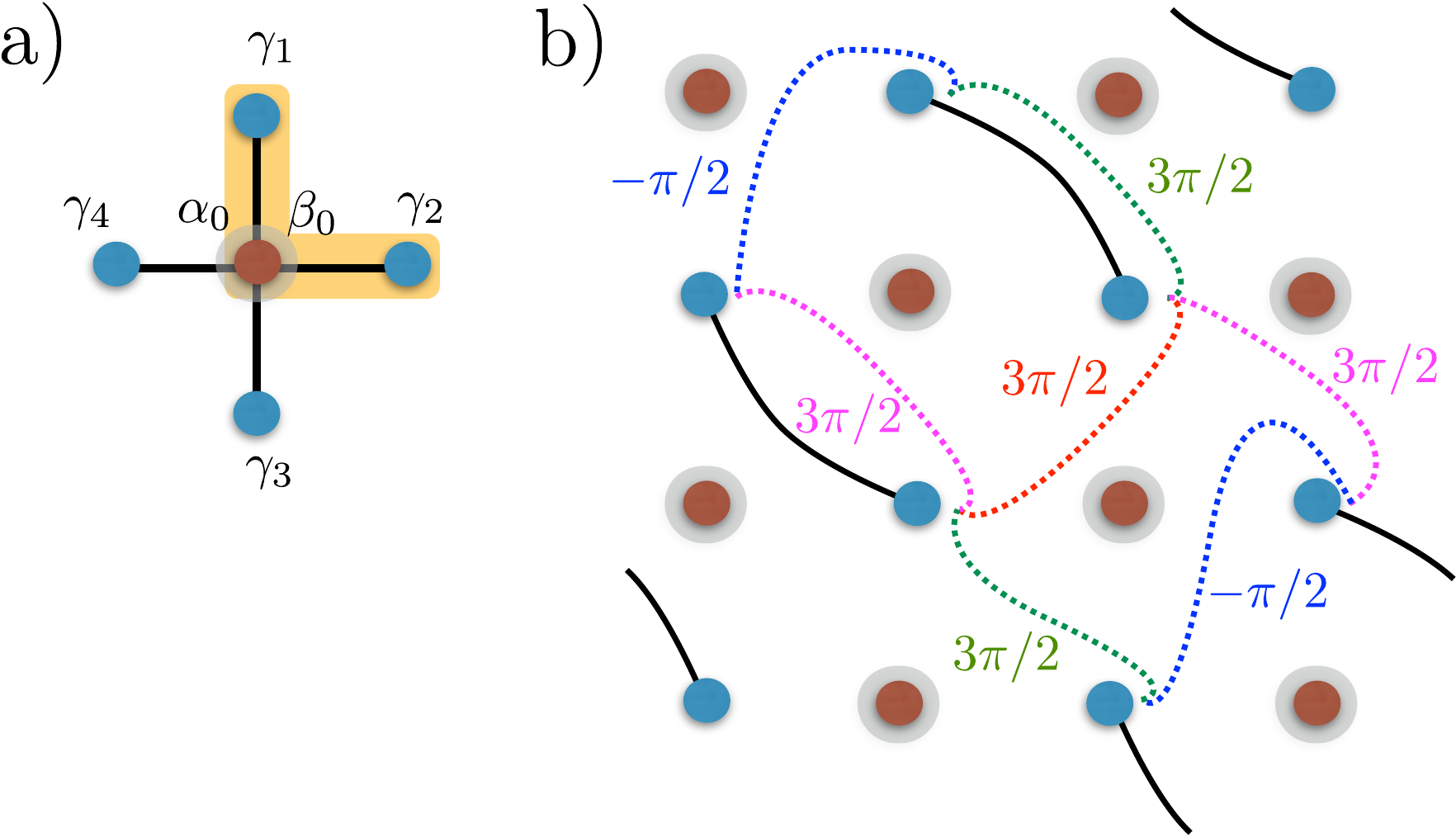}
\caption{Modified square lattice structure {\bf a} A site with the doubly quantized vortex surrounded by singly quantized vortices. The dominant type of interaction within such a node is shaded in yellow. {\bf b} Phase difference structure for a choice of gauge in the modified square lattice.
}\label{fig5}
\end{figure}

Consider first the case  $R_c<\xi$. The dominant interaction is the $g'$ term  as Coulomb interaction on long lengthscale decays faster than the overlap of the Majorana wavefunctions. This interaction term  dominates as it does not involve the smallness due to the screening of the Coulomb interaction, only due to the decay of the Majorana wavefunctions. Following the observation of the previous section that $\rho_{ij} \propto \sin ((\theta_i - \theta_j)/2)$, we see that the interaction is proportional to $ \sin((\theta_{\bj + {\bm \delta}_1} - \theta_{\bj + {\bm \delta}_2})/2)$. This is the interaction of the form $\rho_{\alpha \beta} \rho_{\gamma \gamma}$. The rest of the terms in \eqref{abel17} are canceling each other, since $\alpha_\bj$ and $\beta_\bj$ are different by as if they had a phase difference $\pi$. 

As we noted in the main text, the products $i\alpha_\bj \beta_\bj$ commute with the Hamiltonian and with each other. Thus they are conserved quantities $s_\bj=\pm 1$ signaling the occupation of the Andreev states $c_\bj={1\over 2}(\alpha_\bj + i \beta_\bj)$. This means that we can trace out these degrees of freedom from the model and obtain the hopping amplitudes between the single-vortex sites. The relevant  phase differences $\theta$ are depicted in Fig.\ \ref{fig5}b. We consider for illustration two possible configurations of $s_\bj$: ferromagnetic (all $+1$ or all $-1$) and antiferromagnetic (staggered on the two sublattices). It is easy to see that for the FM configuration the hopping amplitudes on a given bond contributed by the two adjacent double vortex sites add up while for the AF configuration they cancel.  Therefore, in the AF case
the resulting hopping model produces a completely flat Majorana band. Meanwhile for the FM configuration the hopping model will be of the form indicated in Eq.\ (5) of the main text with the nearest neighbor hopping $t=2g'$.  The energy spectrum then consists of a  pair of dispersing bands with energies
\begin{equation}\label{tight}
E_\bk=\pm 4g'\sqrt{\sin^2\left({k_x+k_y\over 2}\right)+\sin^2\left({k_x-k_y\over 2}\right)} 
\end{equation}
where $\bk$ ranges over the reduced Brillouin zone. Occupying the negative energy states in Eq.\ (\ref{tight}) clearly produces lower ground state energy than occupying a flat band at zero energy, therefore hinting that the FM state is the ground state of the system. To prove this we should also consider all other possible occupations of the lattice. Our numerics in the systems up to $6\times 6$ unit cells shows that the FM state is the stable ground state of the system. There is no reason this should change in larger systems. For the screened Coulomb interaction, therefore, a gapless metallic phase with the excitation spectrum (\ref{tight}) is produced.

Now consider the case  $R_c>\xi$. Here the dominant interaction is between the double vortices. If the interaction is the usual Coulomb repulsion, $g\gg g' >0$, then the preferred occupation $s_\bj$ of the double vortices is antiferromagnetic and the hopping model obtained is  the flat Majorana band, as discussed above. Smaller terms involving four single vortex sites can split this degeneracy, but the model thus obtained is not integrable. If the interaction is attractive, $g<0$, $|g|\gg |g'|$, then the preferred occupation of the double vortices is ferromagnetic and the resulting model is the same as for the screened Coulomb, a gapless dispersing Majorana band Eq.\ (\ref{tight}).

\section{2D single Majorana vortex lattice}\label{single ED}

 The exact diagonalization study of the system  on the simple square lattice is performed by transforming the Hamiltonian (\ref{sq2})  to the fermionic basis,  $\alpha_\bj=c^\dagger_\bj+ c^{}_\bj$, $\beta_\bj=i(c^\dagger_\bj-c_\bj^{})$.
The Hamiltonian then becomes
\begin{align}
{\cal H}_{\rm int} &= -g_1 \sum_\bj (2N_\bj-1) (2N_{\bj+x}-1) \nonumber \\
&+g_2\sum_\bj (c^\dagger_\bj -c_\bj )(c^\dagger_{\bj+x}-c^{}_{\bj+x}) \nonumber \\
&\times (c^\dagger_{\bj-y}+c^{}_{\bj-y})(c^\dagger_{\bj-y+x}+c^{}_{\bj-y+x}) \label{fermionic basis}
\end{align}
where $N_\bj= c^\dagger_\bj c^{}_\bj$ denotes the number operator and
 $\bj$ indicates the 2D coordinate $(n,m)$ of the unit cell. If we were to directly diagonalize the many-body Hamiltonian, only a small system can be numerically treated. Fortunately,  ${\cal H}_{\rm int}$ can be block-diagonalized by defining the fermion parity operators
\begin{align}
\hat{F}^x_{n}=(-1)^{\sum_m N_{n,m}},\quad \hat{F}^y_m=(-1)^{\sum_n N_{n,m}},
\end{align}
which commute with the Hamiltonian ${\cal H}_{\rm int}$ and among themselves. Their eigenvalues ($\pm 1$)  are good quantum numbers and label the different blocks of the Hamiltonian. However, these operators are not independent since they are connected by the total fermionic parity operator 
$
\hat{F}=\prod_n \hat{F}^x_{n} =\prod_m \hat{F}^y_{m}.
$

We consider separately the cases when the number $N_x$ of unit cells in the $x$ direction is odd and even. 
The Hamiltonian can be easily block-diagonalized  by $F^x_n=\pm 1$ and $F^y_m=\pm 1$. We are able to numerically solve the block-diagonalized Hamiltonian for a system containing $N_x \times 4 $ unit cells with $N_x$  up to 19 as follows. We first find one of the degenerate ground states $\ket{G}$ in the parity sector $F^x_{n}=1$ and $F^y_m=1$ for all $n$ and $m$. 
We then use the operator $\hat{A}_{\tilde{m}}=\prod_n \alpha_{n,\tilde{m}}$ to generate the remaining ground states. Note that $\hat{A}_{\tilde{m}}$ commutes with ${\cal H}_{\rm int}$ but anticommutes with all $\hat{F}^x_{\tilde{n}}$. When it acts on a ground state it thus flips the sign of all $F^x_{\tilde{n}}$  generating a new ground state in a different parity sector. When we subsequently apply $\hat{A}_{\tilde{m}'}$ with $\tilde{m}'\neq \tilde{m}$ to this new ground state all $F^x_{\tilde{n}}$ flip back. This construction indicates that there exist \emph{at least} two degenerate ground states. For even $N_x$, our numerical results support the two-fold ground state degeneracy. For odd $N_x$, $\hat{A}_{\tilde{m}}$ also flips the sign of $F^y_{\tilde{m}}$. 
Since the number of $\hat{F}^y_{\tilde{m}}$ operators is $N_y$ and $F^y_{\tilde{m}}=\pm 1$ the degenerate ground states are given by
\bee
\ket{F^y_{\tilde{m}_\pm}=\pm 1}=\prod_{\tilde{m}_-}\hat{A}_{\tilde{m}_-}\ket{G}.
\ee
It follows that the number of the degenerate ground states is \emph{at least} $2^{N_y}$. This agrees with the degeneracy that occurs in the extreme anisotropy limit $g_2=0$, already discussed in the main text. 

	Interestingly, the systems with even and odd $N_x$ exhibit different physical properties even in the thermodynamic limit. When $N_x$ is even, by performing a $Z_2$ gauge transformation $\alpha_{2l,m} \rightarrow -\alpha_{2l,m}$, the Hamiltonian ${\cal H}_{\rm int}$ changes the sign. That is, when the many-body state has energy $E$, the state after the gauge transformation has energy $-E$. This many-body version of the particle-hole symmetry shows that $g_1,g_2\geq 0$ describes identical physics as  $g_1,g_2\leq 0$. However, for odd $N_x$, $\alpha_{2l,m} \rightarrow -\alpha_{2l,m}$ does not simply flip the sign of  ${\cal H}_{\rm int}$ due to the frustration at the boundary with the periodic boundary condition. Hence, systems with positive $g_1$ and $g_2$ are different from those with negative $g_1$ and $g_2$ in this case.

	After obtaining the many-body wavefunctions of the ground states from the exact diagonalization, the order parameter $\Delta_1=-i\langle\alpha_\bj\beta_\bj\rangle$ can be computed as a ground state expectation value  in different parity sectors. We mainly focus on odd $N_x$. Because $\hat{A}_{\tilde{m}}=\prod_n \alpha_{n,\tilde{m}}$ connects the ground states in the different parity sectors,  $\Delta_1$ must be computed in only one of the parity sectors, say $F^x_{\tilde{n}}=1$ and $F^y_{\tilde{m}}=1$ for all $\tilde{n}$ and $\tilde{m}$, as shown in Fig.\ \ref{fig3}d. The expectation value flips the sign when we consider the ground state with  parity $F^y_m=-1$.





\begin{thebibliography}{10}
\bibitem{cook2} Ref.\ \cite{cook1} considers Majorana zero mode at the end of a thin topological insulator wire. The calculation for a vortex in a 2D STI surface is similar and leads to the same result regarding the higher energy bound states.
\end{thebibliography}
\end{document}